\documentclass[twocolumn,prl,aps,superscriptaddress,showpacs]{revtex4}

\usepackage{bm}\usepackage{graphicx}

\begin{document}

\title{Understanding the complex phase diagram of uranium: the role of electron-phonon coupling}

\author{S. Raymond}
\affiliation{SPSMS, UMR-E 9001, CEA-INAC/UJF-Grenoble 1, 38054 Grenoble, France}
\author{J. Bouchet}
\affiliation{CEA, DAM, DIF, F-91297 Arpajon, France}
\author{G.H. Lander}
\affiliation{European Commission, Joint Research Centre, Institute for Transuranium Elements, Postfach 2340, D-76125 Karlsruhe, Germany}
\author{M. Le Tacon}
\affiliation{European Synchrotron Radiation Facility, Boite postale 220X, F-38043 Grenoble,France}
\affiliation{Max-Planck-Institut f\"ur Festk\"orperforschung, Heisenbergstra\ss e 1, D-70569 Stuttgart, Germany}
\author{G. Garbarino}
\affiliation{European Synchrotron Radiation Facility, Boite postale 220X, F-38043 Grenoble,France}
\author{M. Hoesch}
\affiliation{European Synchrotron Radiation Facility, Boite postale 220X, F-38043 Grenoble,France}
\author{J.-P. Rueff}
\affiliation{Synchrotron SOLEIL, L'Orme des Merisiers, Saint Aubin, BP 48, 91192 Gif-sur-Yvette Cedex, France}
\author{M. Krisch}
\affiliation{European Synchrotron Radiation Facility, Boite postale 220X, F-38043 Grenoble,France}
\author{J.C. Lashley}
\affiliation{Los Alamos National Laboratory, New Mexico 87545, USA}
\author{R.K. Schulze}
\affiliation{Los Alamos National Laboratory, New Mexico 87545, USA}
\author{R.C. Albers}
\affiliation{Los Alamos National Laboratory, New Mexico 87545, USA}

\date{\today}

\begin{abstract}
We report an experimental determination of the dispersion of the soft phonon mode along [1,0,0] in uranium as a function of pressure. The energies of these phonons increase rapidly, with conventional behavior found by 20 GPa, as predicted by recent theory. New calculations demonstrate the strong pressure (and momentum) dependence of the electron-phonon coupling, whereas the Fermi-surface nesting is surprisingly independent of pressure. This allows a full understanding of the complex phase diagram of uranium, and the interplay between the charge-density wave and superconductivity. 
\end{abstract}


\maketitle

Competition between different ground states is a central issue in condensed-matter physics. Much discussed examples are those between superconductivity and magnetism in cuprates, iron pnictides, and heavy fermions. Equally important is that between the Charge Density Wave (CDW) and superconductivity, as shown by recent progress reported for the transition-metal dichalcogenides \cite{Morosan} and elements under pressure \cite{Degtyareva}. The present paper sheds new light on the mechanisms that govern such an interplay between CDW and superconductivity in uranium. 

At ambient pressure, uranium is the only element to exhibit a phase transition to a CDW state below  $T_{0}$ = 43 K \cite{Lander,chromium}. This transition has been ascribed to nesting of certain features of the Fermi surface \cite{Fast}.  The ambient pressure superconductivity of uranium, reported in the early studies \cite{Lander} below 1 K, is still controversial \cite{Graf} but it is agreed that  the superconducting temperature $T_{c}$ exhibits a maximum of about 2 K at around 1.5 GPa, when the CDW disappears \cite{Lander}.

The room temperature crystal structure of uranium is of much interest, since it is unique for an element at ambient pressure. Uranium exists in an unusual orthorhombic structure ($\alpha$-U phase, space group Cmcm) \cite{Jacob} and adopts this structure to at least 100 GPa \cite{Bihan}. Similar orthorhombic structures are found in Ce \cite{Ellinger}, Am and Cm \cite{Heathman2} at higher pressures, and are understood as a consequence of the $f$ electrons in these materials being squeezed into itinerant states at high pressure. The key aspect, which stabilises the low-symmetry orthorhombic $\alpha$-U structure \cite{Soderlind}, is the narrow band (2-3 eV wide) containing about three 5$f$ electrons at the Fermi level. At low temperature, the CDW state, to a first approximation may be considered as a doubling of the $a$-axis of the unit cell and this structure is called $\alpha_{1}$-U.  

Recent progress in band structure calculations allows the accurate determination of the phonon spectrum of actinide-based materials \cite{Wong,Raymond}. 
Treating the 5$f$ electrons as itinerant, the unusual phonon spectrum of $\alpha$-U was reproduced only in 2008 by {\it ab initio\/} calculations \cite{Bouchet}, almost 30 years after its experimental determination \cite{Crummet}. 
Importantly, these calculations incorporate all 5$f$ electrons as itinerant in the correct orthorhombic $\alpha$-U structure.  If the number of 5$f$ states is varied, or the incorrect crystal structure used, the soft phonon in the $\Sigma_4$ branch is {\it not\/} reproduced \cite{Bouchet}. A prediction of this calculation is that under pressure the energy of the soft phonon with $\Sigma_{4}$ symmetry \cite{Crummet} in the [100] direction {\it increases\/}, until the anomaly disappears near 20 GPa. In the present paper, we report Inelastic X-ray Scattering (IXS) data that confirm the major changes predicted in the phonon spectra on applying pressure; thus benchmarking the theory. The quantitative agreement between experiment and theory encourages us to perform new calculations aiming to understand the complex phase diagram of uranium at low temperature.  As a function of pressure, the increase in energy of the soft mode is tied directly to changes in the electron-phonon (e-ph) coupling, whereas, surprisingly, the Fermi surface nesting remains unaltered. 

The IXS experiments were performed on a single crystal sample at the beamline ID28 at the European Synchrotron Radiation Facility in Grenoble, France, and the theoretical calculations were performed using density functional theory. Details of the experimental and theoretical methods are given in Supplementary information I and II. Figure 1 shows IXS spectra for different pressures around the key positions in reciprocal space. In this paper we indicate the momentum transfer $\textbf{Q}$ = $\tau$ + ($h$, $k$, $l$), where $\tau$ is a reciprocal lattice vector. It is known from earlier studies \cite{Crummet,Lander}, as well as theory \cite{Bouchet}, that the $\Sigma_{4}$ phonon energies are not strongly dependent on the coordinate $l$ along the [001] axis; thus we are interested primarily in the $h$ parameter along [100]. The data shown in Fig. 1 establish the hardening (increase of energy) of the $\Sigma_4$ mode, as pressure is increased. Figure 2(a) shows the dispersion along the important [100] direction for three pressures for data taken at $\textbf{q}=[h, 0, l]$ with $0 \leq l \leq 0.2$. The minimum is known to occur for $h$ $\approx$ 0.5 \cite{Crummet,Lander}. Figure 2(b) shows the soft-mode energy as a function of pressure. Our data confirm the disappearance of the soft mode under pressure as predicted by Bouchet \cite{Bouchet} (Supplementary information III).  When the pressure was released from 20 to 6 GPa, the soft mode was found reversible with pressure (See Fig. 2b).
\begin{figure}
\centering
\includegraphics[width=10cm]{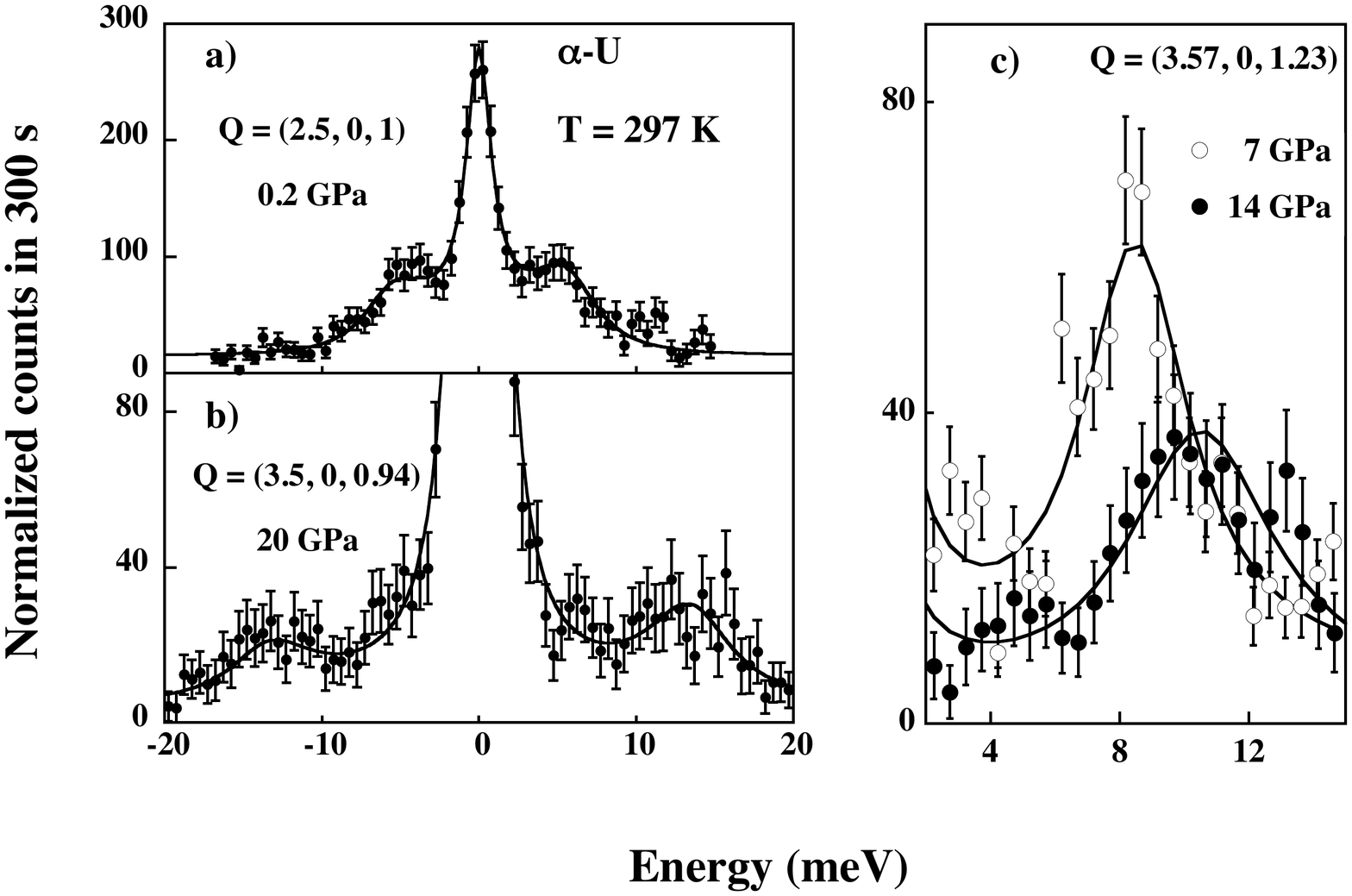}
\vspace{-1.3cm}
\caption{IXS data at ambient temperature taken for $\textbf{q}$ $\approx$ (0.5, 0, 0). Data show both the energy-loss (Stokes) and energy-gain (anti-Stokes) response, as well as the central elastic line at the lowest a) and highest b) pressure measured. c) Zoom on the Stokes peak for intermediate pressures. Note the increase of the phonon energy as the pressure is increased.}
\end{figure}

\begin{figure}
\vspace{-0.7cm}
\centering
\includegraphics[width=12cm]{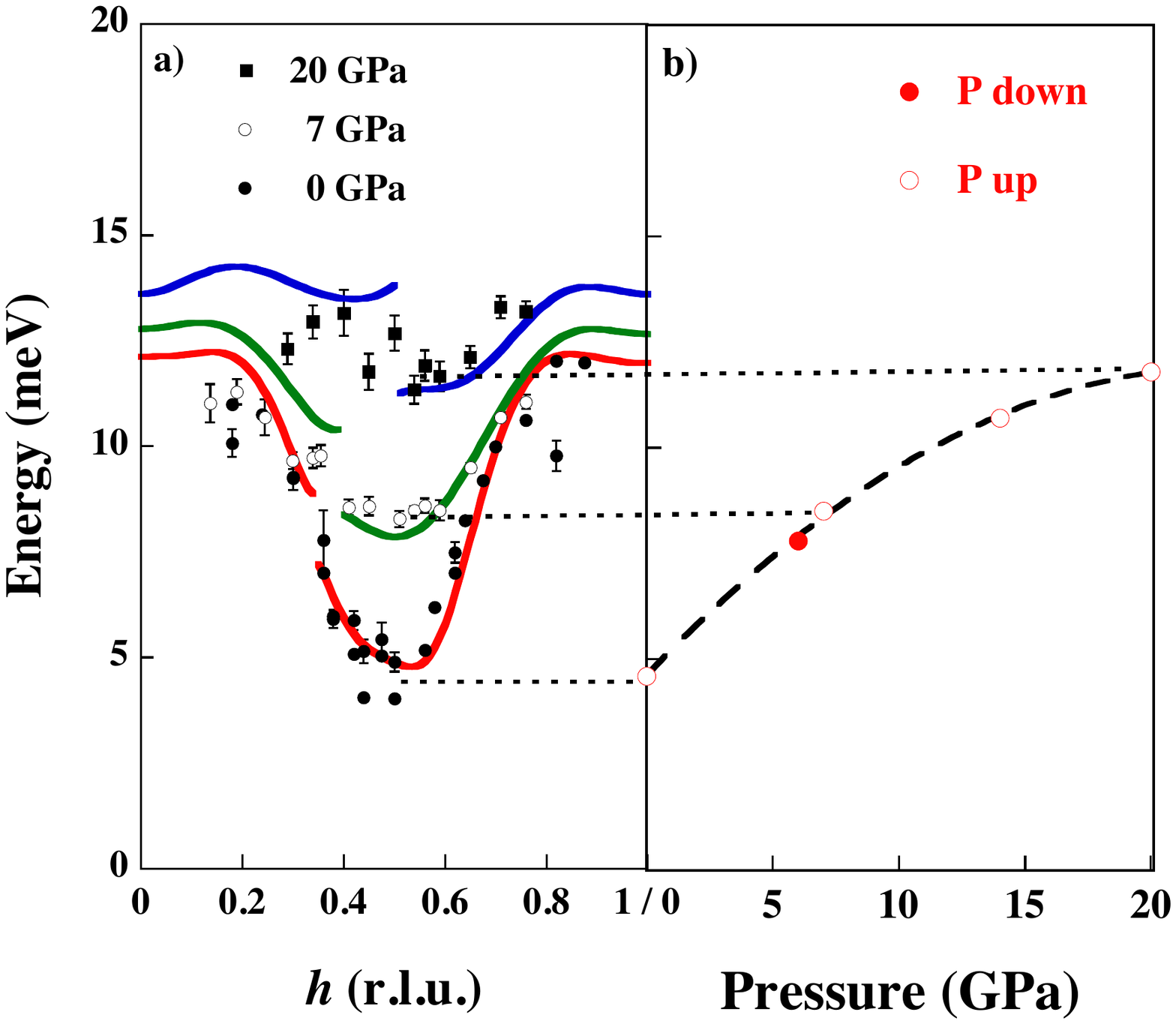}
\vspace{-1.3cm}
\caption{(Color Online) Phonon Energy (a) Experimental (data points with error bars) and theoretical (solid lines) dispersion of the $\Sigma_{4}$ optic component along [h 0 0]. (b) Soft-mode energy as a function of pressure (Lines are guides for the eyes).}
\end{figure}
So far theory and experiments have referred to the orthorhombic $\alpha$-U structure. The excellent agreement between theory and experiment prompted a further examination of the pressure and temperature dependence of parameters that define this structure and that of the CDW ($\alpha_{1}$-U). We calculated the complete band structure for both $\alpha$-U and $\alpha_1$-U, their phonon dispersion curves, and, importantly, the $\textbf{q}$-dependent e-ph coupling $\lambda_{\textbf{q},\nu}$ with
\begin{equation}
\lambda_{\textbf{q},\nu}=\frac{2\gamma_{\textbf{q},\nu}}{\pi N(0)(\hbar \omega_{\textbf{q},\nu})^{2}}
\end{equation}
where $N(0)$ is the density of states at the Fermi level, $\hbar \omega_{\textbf{q},\nu}$ is the phonon energy at wave vector $\textbf{q}$ and phonon mode index $\nu$, and $\gamma_{\textbf{q},\nu}$ is the mode-resolved linewidth (in energy units) resulting from e-ph coupling
 \begin{equation}
\gamma_{\textbf{q},\nu}=2\pi \hbar \omega_{\textbf{q},\nu}\sum_{\textbf{k}}|M_{\textbf{k}+\textbf{q},\textbf{k}}^{\nu}|^{2}\delta(\epsilon_{\textbf{k}})\delta(\epsilon_{\textbf{k}+\textbf{q}})
 \end{equation}
 where $M_{\textbf{k}+\textbf{q},\textbf{k}}^{\nu}$ is the e-ph matrix element,   $\epsilon_{\textbf{k}}$ are the electron eigenvalues, and the sum runs over the Brillouin zone \cite{Liu, Savrasov}.

By examining the phonons and the e-ph coupling, we extract new details about the phase diagram. The key parameters for the $\Sigma_{4}$ mode in the $\alpha$-U structure as a function of momentum transfer along the [100] direction, and for different pressures, are shown in Fig. 3. Figure 3(a) shows the phonon dispersion, Fig. 3(b) shows the e-ph coupling specific to this mode $\lambda_{\textbf{q},\nu}$, and Fig. 3(c) shows the phonon linewidth due to e-ph interaction. The strong dependence along [100] implies that the major difference of the $\alpha_{1}$-U (CDW) from the stable $\alpha$-U structure will be along this direction. The total energy calculations (at $T$=0 K) show that the ground state at ambient pressure is the $\alpha_{1}$-U (CDW), and the cross over to the normal $\alpha$-U structure is calculated to be at just over 1 GPa. When fully relaxing the $\alpha_{1}$-U structure, small components are also found theoretically along [010] and [001], in agreement with experiment \cite{Lander}.
\begin{figure}
\vspace{-2cm}
\centering
\includegraphics[width=18cm]{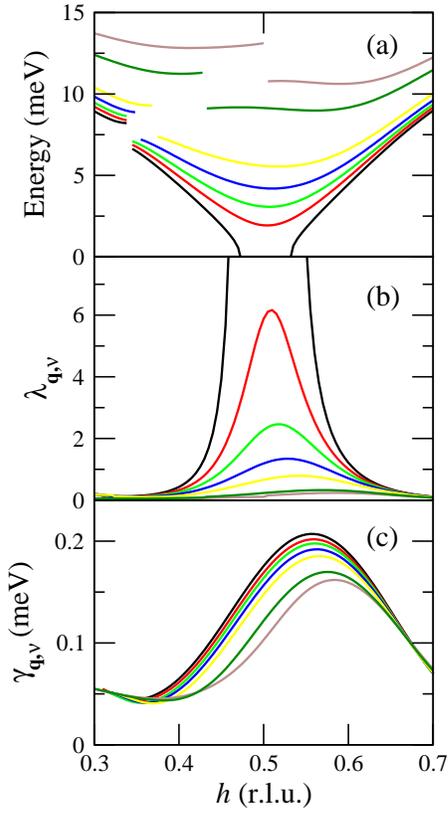}
\vspace{-1cm}
\caption{(Color Online) Results of calculations at $T$=0 K using the experimental volume (See Supplementary information II) (a) Energy of the $\Sigma_4$ phonon mode as a function of the momentum transfer, $\textbf{q}$ = [h, 0, 0]. (b) The e-ph coupling, $\lambda_{\textbf{q},\nu}$. (c) The phonon linewidth due to e-ph interaction, $\gamma_{\textbf{q},\nu}$. The colors represent different pressures in GPa: black 0, red 0.8, light green 2.5, blue 5.1, yellow 10.3, dark green 16, brown 25.}
\end{figure}
The parameter $\gamma_{\textbf{q},\nu}$ is closely connected to the nesting features of the Fermi surface. Clearly, these reach a maximum near the middle of the Brillouin zone along [100]; however, in contrast to $\lambda_{\textbf{q},\nu}$, they do not depend strongly on pressure. We have investigated the Fermi-surface nesting further, as shown in Fig. 4(a) at ambient pressure and in Fig.4(b) at 20 GPa. First, the Fermi surface at ambient pressure closely resembles Fig. 4 of Ref.\cite{Fast} with a nesting vector of magnitude $k_{x}$ $\approx$ 0.5(2$\pi$/$a$) (Supplementary information IV). Second,  comparing Fig.4(a) and Fig.4(b), there is essentially no change as a function of pressure. This central result, that we confirmed with two independent calculations, is unexpected. Thus, the statement that the ÒCDW in uranium is a consequence of the nesting of the Fermi surfaceÓ is incorrect, as the CDW is rapidly destroyed by pressure, disappearing by $\approx$ 1.5 GPa \cite{Lander}, whereas the nesting does not change. 

Indeed, Fermi-surface nesting is present, but although necessary, this is not the unique ingredient for the formation of the $\alpha_{1}$-U (CDW) state, otherwise this phase would not be so sensitive to pressure. The crucial ingredient is the e-ph coupling which allows transferring the energy gain from nesting in the CDW state to the lattice. At high pressure the CDW will not develop, as the e-ph coupling is too weak to transmit the electronic information to the lattice. These conclusions are consistent with the arguments of Johannes and Mazin \cite{Johannes}, who stress that in most cases such phase transitions cannot be ascribed to Fermi-surface nesting alone.
\begin{figure}
\includegraphics[width=9cm,angle=-90]{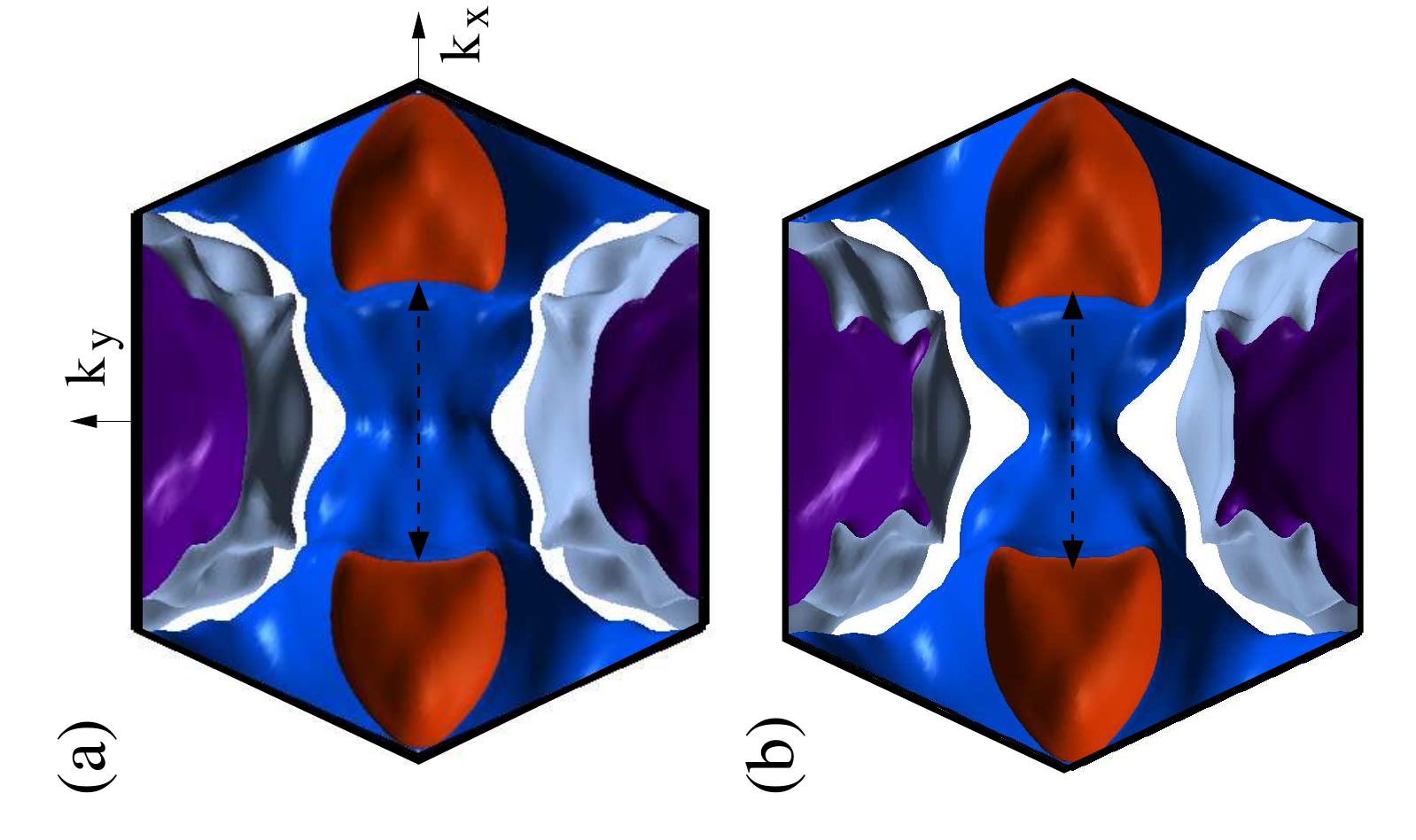}
\caption{(Color Online) The Fermi-surface topology for the $\alpha$-U structure calculated at (a) ambient pressure and (b) at 20 GPa. In each case the diagram shows the [100] and [010] axes at a fixed position $z$ = 1/2 along [001]. The different colors correspond to the different sheets of the Fermi surface. The arrow indicating the nesting vector ($\approx$ 1/2(2$\pi$/$a$) along [100]) has been drawn the same length in both figures.}
\end{figure}
Our calculations for the phonons in the distorted $\alpha_{1}$-U structure (which contains 8 atoms in the unit cell, rather than the 4 in the $\alpha$-U structure) show that the choice for the displacements made previously \cite{Fast} does not result in finite energies for the $\Sigma_{4}$ mode phonons. To correct this we need to preserve the C-face centering in the distorted $\alpha_{1}$-U phase. This gives a lower total energy when compared to the $\alpha_{1}$-U structure adopted in \cite{Fast}.

Finally, we address the fascinating interplay between the CDW and superconductivity, as shown in Fig.5. The open triangles and squares in Fig.5, traced by the solid line, mark the experimental pressure-dependence of $T_{c}$. \cite{Lander}. In order to {\it calculate \/} $T_{c}$, given our knowledge of the e-ph coupling, we use the formula of McMillan \cite{McMillan}:
\begin{equation}
T_{c}=\frac{\theta}{1.45}exp(-\frac{1.04(1+\lambda)}{\lambda-\mu^{*}(1+0.62\lambda)})
\end{equation}
where $\theta$ is the Debye temperature, $\lambda$ is the mass renormalization factor (average e-ph coupling Ð see \cite{Liu,Savrasov}), and $\mu^{*}$ is the e-ph repulsion term. The calculation of $\lambda$ is complex near the transition from the $\alpha$ to $\alpha_1$ phases, due to its large values, and the experimental uncertainties about the nature of the atomic displacements, so it is convenient to anchor the calculations of $T_c$ to the experimental value of $T_{c}$=1.5 K at 3 GPa. This gives a value for $\mu^{*}$ of 0.28. We then deduce [keeping $\mu^{*}$ fixed and using $\lambda_{\textbf{q},\nu}$ from the calculation] $T_c$ in the $\alpha$-U state and these values are shown as red solid points in Fig. 5. Since $\lambda$ strongly reduces with increasing pressure, it is not surprising that $T_c$ also decreases, exactly as found experimentally.
\begin{figure}
\vspace{-1cm}
\includegraphics[width=8cm]{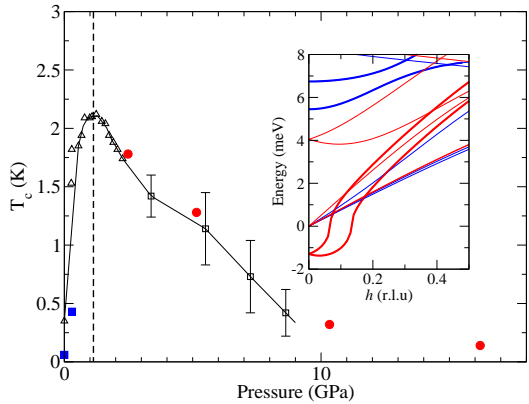}
\caption{(Color Online) $T$-$P$ phase diagram of uranium in the superconducting region. The vertical dashed line (at 1.2 GPa) indicates the calculated transition between the CDW phase (shaded) and normal $\alpha$-U structure in good agreement with experiment. The experimental $T_{c}$ values \cite{Lander} are indicated by open triangles and squares, and the solid line is a guide for the eyes. The calculated values are shown as blue squares ($\alpha_{1}$) and red circles ($\alpha$-U) and were scaled at 3 GPa (see text). The inset shows the phonon dispersion curves in the CDW state ($\alpha_1$ structure) for zero pressure (blue lines) and 5 GPa (red lines). The thicker lines are the branches corresponding to the $\Sigma_{4}$ mode in the $\alpha$-U phase.}
\end{figure}
For $\lambda_{\textbf{q},\nu}$ in the CDW ($\alpha_1$-U) state, on the other hand, the opposite behaviour occurs as a function of pressure. For ambient pressure, the $\Sigma_4$ phonon modes, which are now at the zone center ($\Gamma$) of the new, much smaller, Brillouin zone, have a finite energy, and as the pressure is increased, the phonon modes decrease in energy. The calculations in the inset of Fig. 5 show the low energy phonons in the CDW state at zero and 5 GPa. By the latter pressure, the predicted acoustic phonon energies are negative, showing that the structure is unstable. Thus, initially $\lambda_{\textbf{q},\nu}$ increases with increasing pressure in the CDW state, leading to an increase in $T_c$, as shown by the solid blue squares in Fig. 5, and finally gives the maximum in $T_c$ at the phase transition. Qualitatively, this reproduces satisfactorily the experimentally observed behaviour. 


Our calculations show that the momentum and pressure dependence of the e-ph coupling plays the central role in determining the complex phase diagram of uranium. We are confident of these results; the predictions of the theory about the anomalous phonons are verified by experiment (Fig. 3). However, since the Fermi-surface nesting is independent of pressure (Fig. 4), this alone cannot explain the formation of the CDW \cite{Johannes}. In addition, the theory succeeds in explaining the appearance of the CDW at base temperature and ambient pressure (Fig. 3). It predicts (as observed experimentally) that the CDW will be unstable at 1.2 GPa (Fig. 5). Using the McMillan formula, the pressure dependence of $T_{c}$ is explained (Fig. 5), as well as its absolute value, assuming a reasonable value for the electron-repulsion term and phonon mediated Cooper pairing.

The present joint experimental and theoretical investigation has allowed important progress in understanding the complex phase diagram of uranium, and has shown the crucial role of the narrow 5$f$ band and its influence on the e-ph coupling. Moreover, our study lays the foundation for further extension of theory into strongly correlated systems.


\begin{thebibliography}{99}

\bibitem{Morosan} E. Morosan et al., Nature Physics \textbf{2}, 544 (2006) and references therein.
\bibitem{Degtyareva} O. Degtyareva et al., Phys. Rev. Lett. \textbf{99}, 155505 (2007) and references therein.
\bibitem{Lander} G. H. Lander et al., Advances in Physics \textbf{43}, 1 (1994) and references therein.
\bibitem{chromium} The element chromium exhibits a spin-density wave with an associated CDW, but the latter is not the primary order parameter. Other elements exhibit a CDW at elevated pressure, see M.I. McMahon and R.J. Nelmes, Chem. Soc. Rev. \textbf{35}, 943 (2006) 
\bibitem{Fast} L. Fast et al., Phys. Rev. Lett. \textbf{81}, 2978 (1998)
\bibitem{Graf} See D. Graf et al., Phys. Rev. B \textbf{40}, 241101 (2009)
\bibitem{Jacob} C. W. Jacob and B. E. Warren, J. Am. Chem. Soc. \textbf{59}, 2588 (1937)
\bibitem{Bihan} T. Le Bihan et al., Phys. Rev. B \textbf{67} 134102 (2003)
\bibitem{Ellinger} F.H. Ellinger and W.H. Zachariasen, Phys. Rev. Lett. \textbf{32}, 773 (1974)
\bibitem{Heathman2} S. Heathman et al., Science \textbf{309}, 110 (2005)
\bibitem{Soderlind} P. S\"oderlind et al., Nature \textbf{374}, 524 (1995)
\bibitem{Wong} J. Wong et al., Science \textbf{301}, 1078 (2003) 
\bibitem{Raymond} S. Raymond et al., Phys. Rev. Lett. \textbf{96}, 237003 (2006)
\bibitem{Bouchet} J. Bouchet, Phys. Rev. B \textbf{77}, 024113 (2008)
\bibitem{Crummet} W. P. Crummett et al., Phys. Rev. B \textbf{19}, 6028 (1979)
\bibitem{Liu} A. Y. Liu and A. A. Quong, Phys. Rev. B \textbf{53} R7575 (1996)
\bibitem{Savrasov} S. Y. Savrasov and D. Y. Savrasov, Phys. Rev. B \textbf{54}, 16487 (1996)
\bibitem{Johannes} M. D. Johannes and I. I. Mazin, Phys. Rev. B \textbf{77}, 165135 (2008)
\bibitem{McMillan} W.L. McMillan, Phys. Rev. \textbf{167}, 331 (1968)

\end{thebibliography}
\end{document}